\documentclass[aps,twocolumn,amssymb]{revtex4}
\usepackage{amssymb}
\usepackage{graphicx}
\usepackage{color}
\usepackage{textcomp}
\usepackage{ulem}
\begin{document}

\title{Correlations and fluctuations in a magnetized three-flavor PNJL model with and without inverse magnetic catalysis effect}
\author{Shijun Mao}
 \email{maoshijun@mail.xjtu.edu.cn}
\affiliation{School of Science, Xi'an Jiaotong University, Xi'an, Shaanxi 710049, China}

\begin{abstract}
The correlations $\chi^{BQ}_{11},\ \chi^{BS}_{11} ,\ \chi^{QS}_{11}$ and quadratic (quartic) fluctuations $\chi^B_{2,4},\ \chi^Q_{2,4},\ \chi^S_{2,4}$ of baryon number $B$, electric charge $Q$ and strangeness $S$ are investigated in a three-flavor Polyakov loop extended Nambu-Jona-Lasinio model at finite temperature and magnetic field. The inverse magnetic catalysis (IMC) effect is introduced through the magnetic field dependent parameters $G(eB)$ or $T_0(eB)$, and we make comparison of the results in the cases with and without IMC effect. Since including IMC effect does not change the strength of phase transition under external magnetic field, it does not lead to qualitative difference in the correlations and fluctuations, but modifies their values. Under vanishing and nonvanishing magnetic field, the correlations and fluctuations increase with temperature, and then show the peak around the pseudocritical temperatures of chiral restoration and deconfinement phase transitions. The peak structure in $\chi^{BQ}_{11}$, $\chi^B_{4}$ and $\chi^Q_{4}$ are much more apparent than in other correlations and fluctuations. The correlations and fluctuations along the phase transition line under external magnetic field are characterized by the scaled correlations ${\hat {\chi}}_{11}^{BQ}=\frac{\chi_{11}^{BQ}(eB,T_{pc}^c(eB))}{\chi_{11}^{BQ}(eB=0,T_{pc}^c(eB=0))}$, ${\hat {\chi}}_{11}^{BS}=\frac{\chi_{11}^{BS}(eB,T_{pc}^c(eB))}{\chi_{11}^{BS}(eB=0,T_{pc}^c(eB=0))}$, ${\hat {\chi}}_{11}^{QS}=\frac{\chi_{11}^{QS}(eB,T_{pc}^c(eB))}{\chi_{11}^{QS}(eB=0,T_{pc}^c(eB=0))}$ and scaled quadratic (quartic) fluctuations ${\hat {\chi}}_{2,4}^{B}=\frac{\chi_{2,4}^{B}(eB,T_{pc}^c(eB))}{\chi_{2,4}^{B}(eB=0,T_{pc}^c(eB=0))}$, ${\hat {\chi}}_{2,4}^{Q}=\frac{\chi_{2,4}^{Q}(eB,T_{pc}^c(eB))}{\chi_{2,4}^{Q}(eB=0,T_{pc}^c(eB=0))}$, ${\hat {\chi}}_{2,4}^{S}=\frac{\chi_{2,4}^{S}(eB,T_{pc}^c(eB))}{\chi_{2,4}^{S}(eB=0,T_{pc}^c(eB=0))}$ at the pseudocritical temperature $T_{pc}^c$ of chiral restoration phase transition. They increase with magnetic fields due to the increase of phase transition strength under magnetic fields. Among them, ${\hat {\chi}}_{11}^{BQ}$ increases fastest, which may serve as the magnetometer of QCD.
\end{abstract}

\date{\today}

\maketitle

\section{Introduction}
Motivated by the strong magnetic field in the core of compact stars and in the initial stage of relativistic heavy ion collisions, the study on Quantum Chromodynamics (QCD) phase structure under external electromagnetic fields has attracted much attention~\cite{review0,review1,review2,review3,review4,review5,lattice1,lattice2,lattice2sep,lattice4,lattice5,lattice6,lattice7,lattice9,ding2025,lattice8,fukushima,mao,maosep1,maosep2,maosep3,maosep4,kamikado,bf1,bf13,bf2,bf3,bf5,bf51,bf52,bf8,bf9,bf11,db1,db1sep,db2,db3,db5,db6,pnjl1,pnjl1sep,pnjl2,pnjl3,pnjl4,pqm,ferr1,ferr2,mhuang,meimao1,t0effect,meihuangmao,t0effectmao,efield1,efield2,efield3,efield4}. 

The Lattice QCD (LQCD) simulations~\cite{lattice1,lattice2,lattice2sep,lattice4,lattice5,lattice6,lattice7,lattice9,ding2025} observe the inverse magnetic catalysis (IMC) phenomena of $u$ and $d$ quarks, which means the decreasing chiral condensates near the pseudocritical temperature $T^c_{pc}$ of chiral restoration phase transition and the decreasing pseudocritical temperature $T^c_{pc}$ under external magnetic field. Meanwhile, it is reported that the renormalized Polyakov loop increases with magnetic fields and the transition temperature of deconfinement decreases as the magnetic field grows~\cite{lattice1,lattice2,lattice2sep,lattice4,lattice5,lattice9}. On analytical side, how to explain the inverse magnetic catalysis phenomena is the open question. Many scenarios are proposed~\cite{fukushima,mao,maosep1,maosep2,maosep3,maosep4,kamikado,bf1,lattice9,bf13,bf2,bf3,bf5,bf51,bf52,bf8,bf9,bf11,db1,db1sep,db2,db3,db5,db6,pnjl1,pnjl1sep,pnjl2,pnjl3,pnjl4,pqm,ferr1,ferr2,mhuang,meimao1,t0effect,meihuangmao,t0effectmao}, such as magnetic inhibition of mesons, sphalerons, gluon screening effect, weakening of strong coupling, and anomalous magnetic moment. 

The thermodynamical properties of QCD matter are also influenced by the external magnetic field. Among the thermodynamic quantities, the correlations and fluctuations of the conserved charges are accessible in both theoretical calculations and experimental measurements. They can serve as the useful probes to study the QCD phase transitions, such as to identify the critical end point (CEP) of QCD phase diagram in the temperature-baryon chemical potential plane~\cite{xwork24,xwork25,xwork26,xwork27,xwork28,xwork29,xwork30} and to serve as the possible magnetometer of QCD (probing the presence of magnetic fields in QCD matter)~\cite{dingref31,addding,ding2022,dingprl2024,ding2025}. However, they are much less explored at the chiral restoration phase transition (crossover) with finite temperature, magnetic field and vanishing chemical potential. Except for the LQCD calculations~\cite{addding,ding2022,dingprl2024,ding2025}, the analytical investigations at vanishing chemical potential, finite temperature and magnetic field~\cite{dingref31,dingref33,dingref34,dingref35,dingref36,mao2pnjl} have been conducted in frame of the hadron resonance gas model, Polyakov loop extended Nambu-Jona-Lasinio (PNJL) model and Polyakov loop extended chiral quark model. It should be mentioned that with three-flavor quarks, the IMC effect is not well considered~\cite{dingref31,dingref33,dingref34,dingref35,dingref36}.

In our current paper, the correlations $\chi^{BQ}_{11},\ \chi^{BS}_{11} ,\ \chi^{QS}_{11}$ and quadratic (quartic) fluctuations $\chi^B_{2,4},\ \chi^Q_{2,4},\ \chi^S_{2,4}$ of baryon number $B$, electric charge $Q$ and strangeness $S$ are studied in a three-flavor PNJL model at finite temperature and magnetic field. The IMC effect is introduced through the magnetic field dependent coupling between quarks $G(eB)$~\cite{bf8,bf9,mao2pnjl,geb1,su3meson4,mao11} and magnetic field dependent interaction between quarks and Polyakov loop $T_0(eB)$~\cite{t0effectmao,t0effect,pnjl3,mao2pnjl}, respectively. The comparison between the results in the cases with and without IMC effect are made. After the brief introduction, Sec.\ref{2fframe} presents our three-flavor magnetized PNJL model and the definition of correlations $\chi^{BQ}_{11},\ \chi^{BS}_{11} ,\ \chi^{QS}_{11}$ and quadratic (quartic) fluctuations $\chi^B_{2,4},\ \chi^Q_{2,4},\ \chi^S_{2,4}$. Section \ref{results} discusses the numerical results of correlations and fluctuations at finite temperature and magnetic field in the cases without IMC effect and with IMC effect. Finally, we give the summary in Sec.\ref{summary}.


\section{theoretical framework}
\label{2fframe}

The three-flavor PNJL model under external magnetic field is defined with the Lagrangian density~\cite{pnjl5,pnjl6,pnjl7,pnjl8,pnjl9,pnjl10,pnjl12},
\begin{eqnarray}
\mathcal{L}&=&\bar{\psi}\left(i\gamma^{\mu}D_{\mu}-\hat{m}_0\right)\psi+\mathcal{L}_{ 4}+\mathcal{L}_{6}-{\cal U}(\Phi,\bar\Phi),\\
\mathcal{L}_{ 4}&=&G\sum_{\alpha=0}^{8}\left[(\bar{\psi}\lambda_{\alpha}\psi)^2+(\bar{\psi}i\gamma_5\lambda_{\alpha}\psi)^2\right],\nonumber \\
\mathcal{L}_{6}&=&-K\left[\text{det}\bar{\psi}(1+\gamma_5)\psi+\text{det}\bar{\psi}(1-\gamma_5)\psi \right],\nonumber\\
{\cal U}(\Phi,{\bar \Phi}) &=&  T^4 \left[-{b_2(t)\over 2} \bar\Phi\Phi -{b_3\over 6}\left({\bar\Phi}^3+\Phi^3\right)+{b_4\over 4}\left(\bar\Phi\Phi\right)^2\right].\nonumber
\label{lagrangian}
\end{eqnarray}
The covariant derivative $D^\mu=\partial^\mu+i Q A^\mu-i {\cal A}^\mu$ couples quarks to the two external fields, the magnetic field ${\bf B}=\nabla\times{\bf A}$ and the temporal gluon field  ${\cal A}^\mu=\delta^\mu_0 {\cal A}^0$ with ${\cal A}^0=g{\cal A}^0_a \lambda_a/2=-i{\cal A}_4$ in Euclidean space. The gauge coupling $g$ is combined with the SU(3) gauge field ${\cal A}^0_a(x)$ to define ${\cal A}^\mu(x)$, and $\lambda_a$ are the Gell-Mann matrices in color space. We consider magnetic field ${\bf B}=(0, 0, B)$ along the $z$-axis by setting $A_\mu=(0,0,x B,0)$ in Landau gauge, which couples quarks of electric charge $Q=\text{diag}(Q_u,Q_d,Q_s)=\text{diag}(2/3 e,-1/3 e,-1/3 e)$. $\hat{m}_0=\text{diag}(m^u_0,m_0^d,m_0^s)$ is the current quark mass matrix in flavor space. The four-fermion interaction $\mathcal{L}_{4}$ represents the interaction in scalar and pseudo-scalar channels, with Gell-Mann matrices $\lambda_{\alpha},\ \alpha=1,2,...,8$ and $\lambda_0=\sqrt{2/3} \mathbf{I}$ in flavor space. The six-fermion interaction or Kobayashi-Maskawa-'t Hooft term $\mathcal{L}_{6}$ is related to the $U_A(1)$ anomaly~\cite{tHooft1,tHooft2,tHooft3,tHooft4,tHooft5}. The Polyakov potential ${\cal U}(\Phi,\bar\Phi)$ describes deconfinement at finite temperature, where $\Phi$ is the trace of the Polyakov loop $\Phi=\left({\text {Tr}}_c L \right)/N_c$, with $L({\bf x})={\cal P} \text {exp}[i \int^\beta_0 d \tau {\cal A}_4({\bf x},\tau)]= \text {exp}[i \beta {\cal A}_4 ]$ and $\beta=1/T$, the coefficient $b_2(t)=a_0+a_1 t+a_2 t^2+a_3 t^3$ with $t=T_0/T$ is temperature dependent, and the other coefficients $b_3$ and $b_4$ are constants.

It is useful to convert the six-fermion interaction into an effective four-fermion interaction in the mean field approximation, and the Lagrangian density can be rewritten as~\cite{3njlrehberg,3pnjlmei}
	\begin{eqnarray}
		\mathcal{L}&=&\bar{\psi}\left(i\gamma^{\mu}D_{\mu}-\hat{m}_0\right)\psi-{\cal U}(\Phi,\bar\Phi) \\
		&+&\sum_{a=0}^{8}\left[K_a^-\left(\bar{\psi}\lambda^a\psi\right)^2+K_a^+\left(\bar{\psi}i\gamma_5\lambda^a\psi\right)^2\right]\nonumber\\	&+&K_{30}^-\left(\bar{\psi}\lambda^3\psi\right)\left(\bar{\psi}\lambda^0\psi\right)+K_{30}^+\left(\bar{\psi}i\gamma_5\lambda^3\psi\right)\left(\bar{\psi}i\gamma_5\lambda^0\psi\right)\nonumber\\	&+&K_{03}^-\left(\bar{\psi}\lambda^0\psi\right)\left(\bar{\psi}\lambda^3\psi\right)+K_{03}^+\left(\bar{\psi}i\gamma_5\lambda^0\psi\right)\left(\bar{\psi}i\gamma_5\lambda^3\psi\right)\nonumber\\	&+&K_{80}^-\left(\bar{\psi}\lambda^8\psi\right)\left(\bar{\psi}\lambda^0\psi\right)+K_{80}^+\left(\bar{\psi}i\gamma_5\lambda^8\psi\right)\left(\bar{\psi}i\gamma_5\lambda^0\psi\right)\nonumber\\	&+&K_{08}^-\left(\bar{\psi}\lambda^0\psi\right)\left(\bar{\psi}\lambda^8\psi\right)+K_{08}^+\left(\bar{\psi}i\gamma_5\lambda^0\psi\right)\left(\bar{\psi}i\gamma_5\lambda^8\psi\right)\nonumber\\	&+&K_{83}^-\left(\bar{\psi}\lambda^8\psi\right)\left(\bar{\psi}\lambda^3\psi\right)+K_{83}^+\left(\bar{\psi}i\gamma_5\lambda^8\psi\right)\left(\bar{\psi}i\gamma_5\lambda^3\psi\right)\nonumber\\	&+&K_{38}^-\left(\bar{\psi}\lambda^3\psi\right)\left(\bar{\psi}\lambda^8\psi\right)+K_{38}^+\left(\bar{\psi}i\gamma_5\lambda^3\psi\right)\left(\bar{\psi}i\gamma_5\lambda^8\psi\right)\nonumber,
		\label{semilagrangian}
	\end{eqnarray}
	with the effective coupling constants
	\begin{eqnarray}
		\label{constants}
		&&K_0^\pm=G\pm\frac{1}{3}K\left(\sigma_u+\sigma_d+\sigma_s\right),\\
		&&K_1^\pm=K_2^\pm=K_3^\pm=G\mp\frac{1}{2}K\sigma_s,\nonumber\\
		&&K_4^\pm=K_5^\pm=G\mp\frac{1}{2}K\sigma_d,\nonumber\\
		&&K_6^\pm=K_7^\pm=G\mp\frac{1}{2}K\sigma_u,\nonumber\\
		&&K_8^\pm=G\mp\frac{1}{6}K\left(2\sigma_u+2\sigma_d-\sigma_s\right),\nonumber\\
		&&K_{03}^\pm=K_{30}^\pm=\mp\frac{1}{2\sqrt{6}}K\left(\sigma_u-\sigma_d\right),\nonumber\\
		&&K_{08}^\pm=K_{80}^\pm=\mp\frac{\sqrt{2}}{12}K\left(\sigma_u+\sigma_d-2\sigma_s\right),\nonumber\\
		&&K_{38}^\pm=K_{83}^\pm=\pm\frac{1}{2\sqrt{3}}K\left(\sigma_u-\sigma_d\right),	\nonumber
	\end{eqnarray}
	and chiral condensates
	\begin{eqnarray}
		\sigma_u=\langle\bar{u}u\rangle, \  \sigma_d=\langle\bar{d}d\rangle, \  \sigma_s=\langle\bar{s}s\rangle.
	\end{eqnarray}
	
The thermodynamic potential in mean field level contains the mean field part and quark part
	\begin{eqnarray}
		\label{omega1}
		\Omega &=&2 G(\sigma_u^2+\sigma_d^2+\sigma_s^2)-4K\sigma_u\sigma_d\sigma_s+{\cal U}(\Phi,\bar\Phi)+\Omega_q,\nonumber \\
		\Omega_q &=& - \sum_{f=u,d,s}\frac{|Q_f B|}{2\pi}\sum_{l}\alpha_l \int \frac{d p_z}{2\pi} \Bigg[3E_f \nonumber\\
		&+& T\ln\left(1+3\Phi e^{-\beta E_f^+}+3{\bar \Phi}e^{-2\beta E_f^+}+e^{-3\beta E_f^+}\right)\nonumber\\
&+& T\ln\left(1+3{\bar \Phi} e^{-\beta E_f^-}+3{ \Phi}e^{-2\beta E_f^-}+e^{-3\beta E_f^-}\right)\Bigg],\nonumber
	\end{eqnarray}
where $f=u,d,s$ means quark flavors, $l$ Landau levels, $\alpha_l=2-\delta_{l0}$ spin factor and $E_f^\pm=E_f \pm \mu_f$ contains quark energy $E_f=\sqrt{p^2_z+2 l |Q_f B|+m_f^2}$ of longitudinal momentum $p_z$ and effective quark masses $m_u=m_0^u-4G\sigma_u+2K\sigma_d\sigma_s$, $m_d=m_0^d-4G\sigma_d+2K\sigma_u\sigma_s$, $m_s=m_0^s-4G\sigma_s+2K\sigma_u\sigma_d$, and quark chemical potential $\mu_u=\frac{\mu_B}{3}+\frac{2\mu_Q}{3}$, $\mu_d=\frac{\mu_B}{3}-\frac{\mu_Q}{3}$, $\mu_s=\frac{\mu_B}{3}-\frac{\mu_Q}{3}+\frac{\mu_S}{3}$, with $\mu_B,\ \mu_Q\ ,\mu_S$ the chemical potential corresponding to the baryon number $B$, electric charge $Q$ and strangeness $S$, respectively.

The ground state at finite temperature, chemical potential and magnetic field is determined by minimizing the thermodynamic potential
\begin{eqnarray}
\label{gapeqs}
 \frac{\partial \Omega }{\partial \sigma_f}&=&0,\ (f=u,d,s),\nonumber\\
 \frac{\partial \Omega }{\partial \Phi}&=&0,\\
  \frac{\partial \Omega }{\partial {\bar \Phi}}&=&0.\nonumber
\end{eqnarray}
The thermodynamic potential $\Omega$ is a function of order parameters (chiral condensates $\sigma_f$ and Polyakov loop $\Phi, {\bar \Phi}$), and hence we obtain five coupled gap equations.

At vanishing chemical potential and finite temperature, the chiral symmetry restoration and deconfinement process are smooth crossover. By considering the derivative of the order parameters (chiral condensates and Polyakov loop) with respect to the temperature, the pseuocritical temperatures $T_{pc}^c$ and $T_{pc}^d$ of chiral restoration and deconfinement phase transitions are determined by the location of the peak of $\frac{d \sigma_{ud}}{d T}\ \text{with}\  \sigma_{ud}=\frac{\sigma_u+\sigma_d}{2}$ and $\frac{d \Phi}{d T}$, respectively. The strength of chiral restoration and deconfinement phase transitions is characterized by the peak value of $\frac{d \sigma_{ud}}{d T}$ and $\frac{d \Phi}{d T}$, respectively. Note that at vanishing chemical potential, we have $\Phi={\bar \Phi}$.

The fluctuations and correlations of baryon number $B$, electric charge $Q$ and strangeness $S$ can be obtained by taking the derivatives of the thermodynamic potential $\Omega$ with respect to the chemical potentials $\hat{\mu}_X=\mu_X/T,\ (X=B,\ Q,\ S)$, evaluated at zero chemical potential
\begin{eqnarray}
\chi_{i,j}^{B,Q,S}&=&-\frac{\partial^{i+j+k}(\Omega/T^4)}{\partial \hat{\mu}_B^i \partial \hat{\mu}_Q^j \partial \hat{\mu}_S^k}{\bigg |}_{{\mu}_X=0}.
\end{eqnarray}
In this work, we focus on the correlations $\chi^{BQ}_{11}$, $\chi^{BS}_{11}$, $\chi^{QS}_{11}$, quadratic fluctuations $\chi^B_2,\ \chi^Q_2,\ \chi^S_2$ and quartic fluctuations $\chi^B_4,\ \chi^Q_4,\ \chi^S_4$ at finite temperature, magnetic field and vanishing chemical potential.

\begin{figure*}[htb]
\includegraphics[width=7cm]{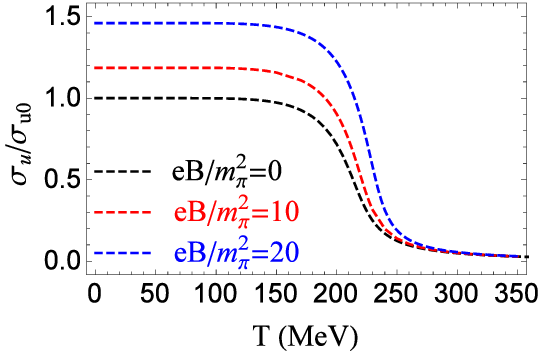}\includegraphics[width=7cm]{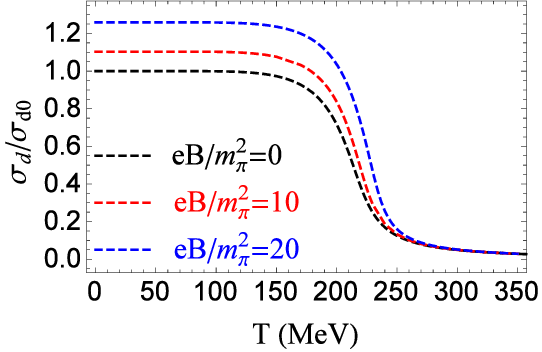}\\
\includegraphics[width=7cm]{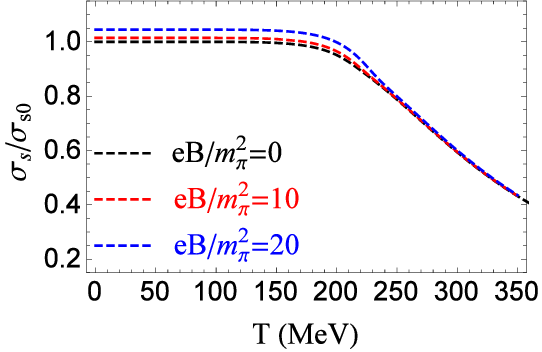}\includegraphics[width=7cm]{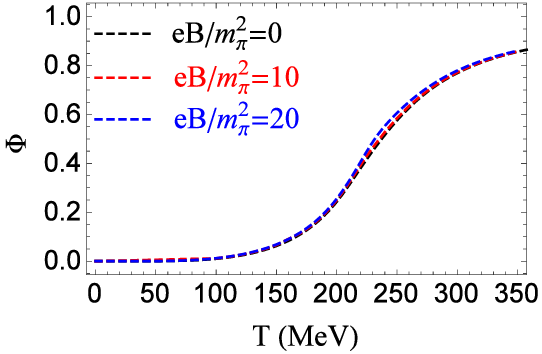}\\
\includegraphics[width=7cm]{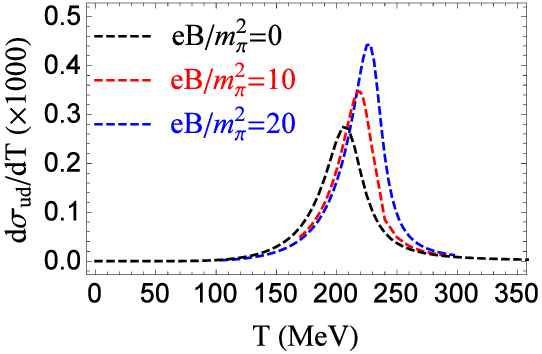}\includegraphics[width=6.7cm]{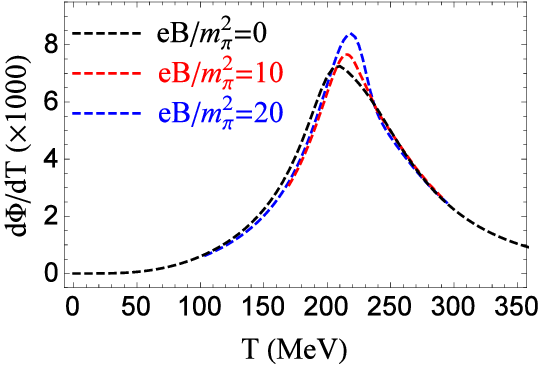}
\caption{(first two rows) The chiral condensates $\sigma_u/\sigma_{u0},\ \sigma_d/\sigma_{d0},\ \sigma_s/\sigma_{s0}$ and Polyakov loop $\Phi$ as functions of temperature with vanishing chemical potential and fixed magnetic field $eB/m^2_\pi=0,\ 10,\ 20$. Here, $\sigma_{u0},\ \sigma_{d0},\ \sigma_{s0}$ means up, down, strange quark chiral condensate in vacuum with vanishing temperature, chemical potential and magnetic field, respectively. (last row) The derivative of chiral condensate and Polyakov loop with respect to the temperature $\frac{d \sigma_{ud}}{d T}$ (left panel), $\frac{d \Phi}{d T}$ (right panel) as functions of temperature at vanishing chemical potential and fixed magnetic field $eB/m^2_\pi=0,\ 10,\ 20$.}
\label{orderparameterg0t0}
\end{figure*}

\begin{figure*}[htb]
\begin{center}
\includegraphics[width=6cm]{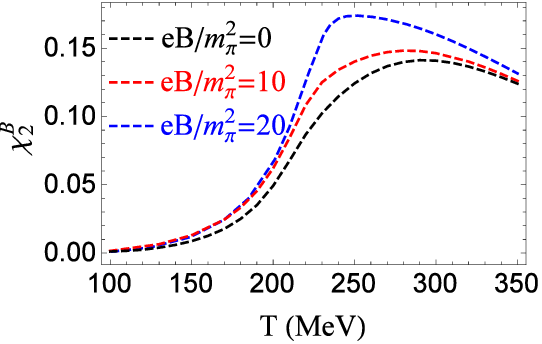}\includegraphics[width=6cm]{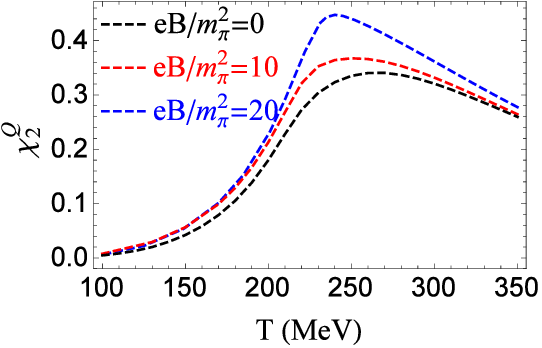}\includegraphics[width=6cm]{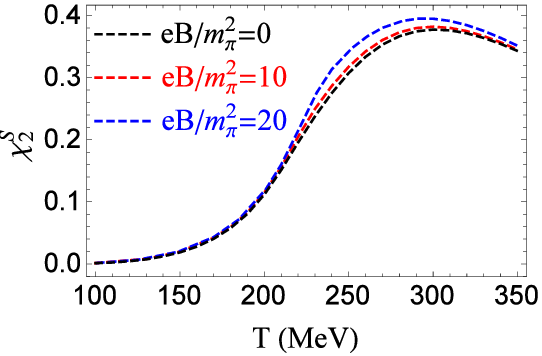}\\
\includegraphics[width=6cm]{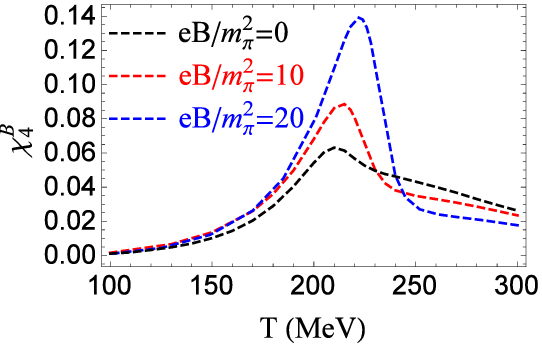}\includegraphics[width=6cm]{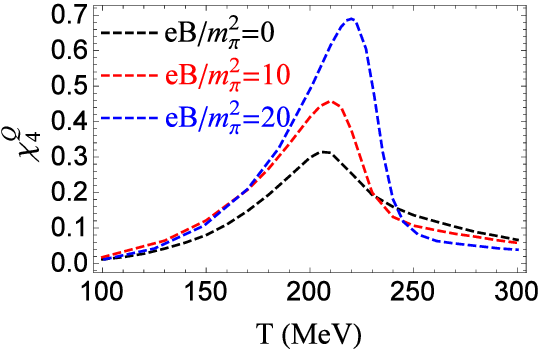}\includegraphics[width=6cm]{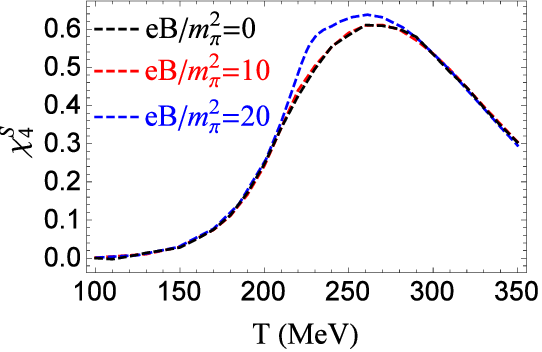}\\
\includegraphics[width=6cm]{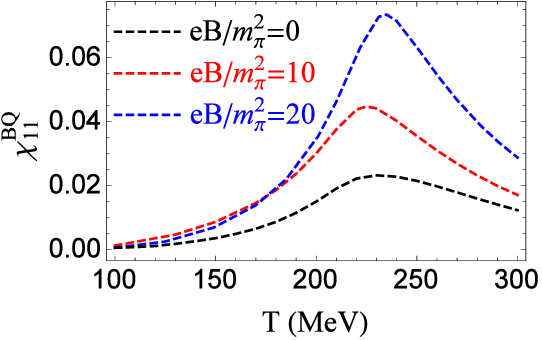}\includegraphics[width=6cm]{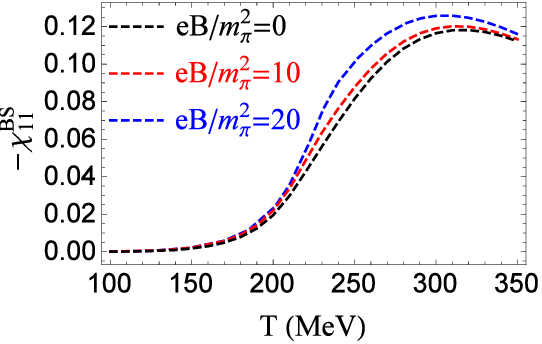}\includegraphics[width=6cm]{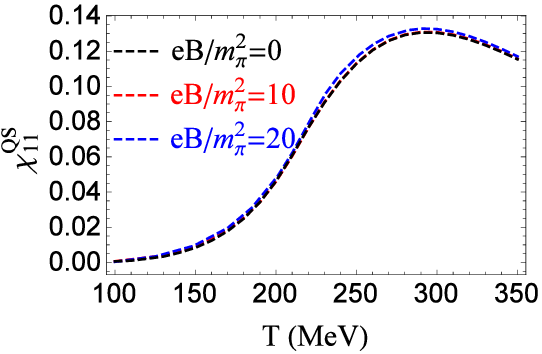}
\end{center}
\caption{Quadratic fluctuations $\chi^B_2,\ \chi^Q_2,\ \chi^S_2$ (first row), quartic fluctuations $\chi^B_4,\ \chi^Q_4,\ \chi^S_4$ (second row) and correlations $\chi^{BQ}_{11},\ -\chi^{BS}_{11},\ \chi^{QS}_{11}$ (last row) as functions of temperature with vanishing chemical potential and fixed magnetic field $eB/m^2_{\pi}=0,\ 10,\ 20$.}
\label{figxng0t0}
\end{figure*}

\begin{figure*}[htb]
\begin{center}
\includegraphics[width=6cm]{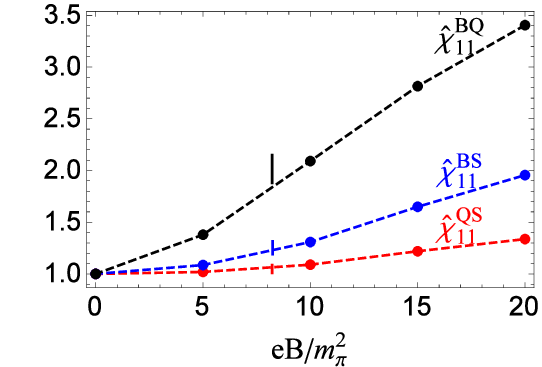}\includegraphics[width=6cm]{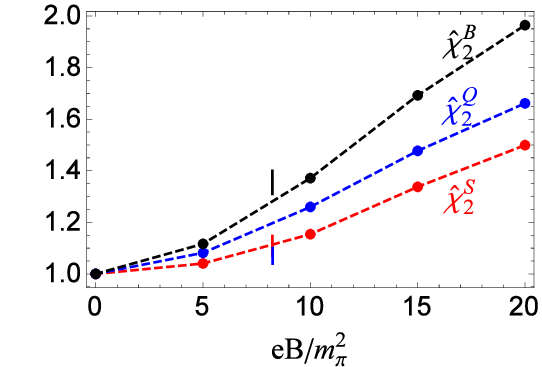}\includegraphics[width=6cm]{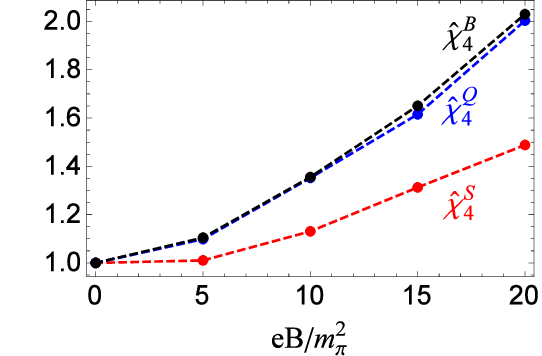}
\end{center}
\caption{The scaled correlations ${\hat\chi}^{BQ}_{11},\ {\hat\chi}^{BS}_{11},\ {\hat\chi}^{QS}_{11}$ (left panel), scaled quadratic fluctuations ${\hat\chi}^B_2,\ {\hat\chi}^Q_2,\ {\hat\chi}^S_2$ (middle panel) and scaled quartic fluctuations ${\hat\chi}^B_4,\ {\hat\chi}^Q_4,\ {\hat\chi}^S_4$ (right panel) at the pseudocritical temperature of chiral restoration phase transition as functions of magnetic field. The vertical lines are the LQCD results~\cite{ding2022,ding2025,dingprl2024}.}
\label{figxntpcg0t0}
\end{figure*}
\begin{figure}[htb]
\begin{center}
\includegraphics[width=7cm]{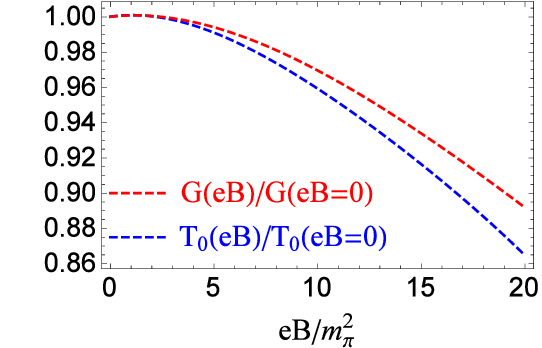}\\
\includegraphics[width=7cm]{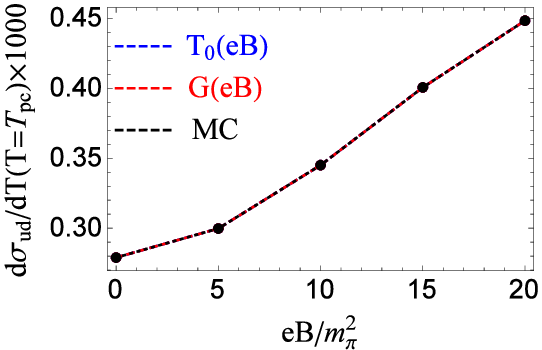}
\end{center}
\caption{(upper panel) Magnetic field dependent parameters $G(eB)$ (red line) and $T_0(eB)$ (blue line) fitted from LQCD reported decreasing pseudocritical temperature of chiral restoration phase transition $T_{pc}^c(eB)/T_{pc}^c(eB=0)$ under external magnetic field~\cite{lattice1}. (lower panel) Strength of chiral restoration phase transition ($\frac{d \sigma_{ud}}{d T}$ at $T=T_{pc}^c$) under external magnetic field with IMC effect (blue and red lines) and without IMC effect (black line).}
\label{figimcparameter}
\end{figure}
\begin{figure*}[htb]
\begin{center}
\includegraphics[width=6cm]{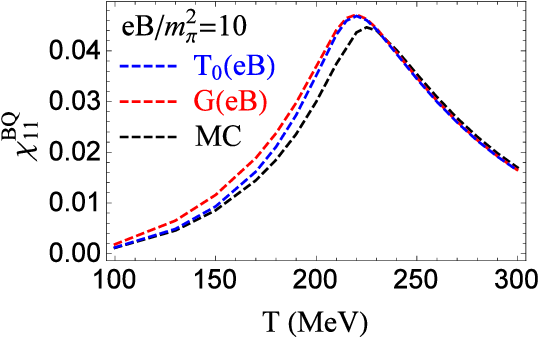}\includegraphics[width=6cm]{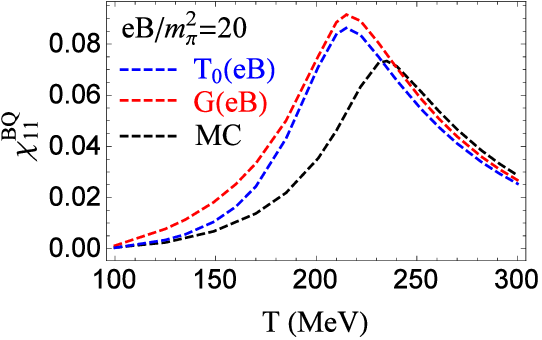}\includegraphics[width=5.8cm]{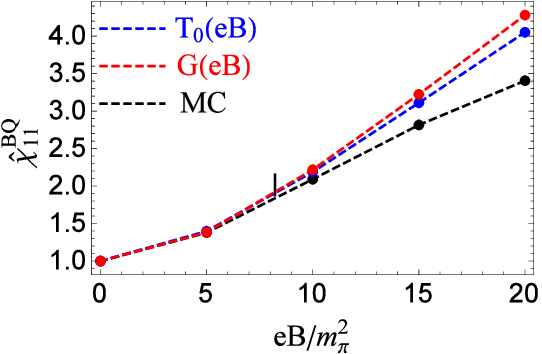}\\
\includegraphics[width=6cm]{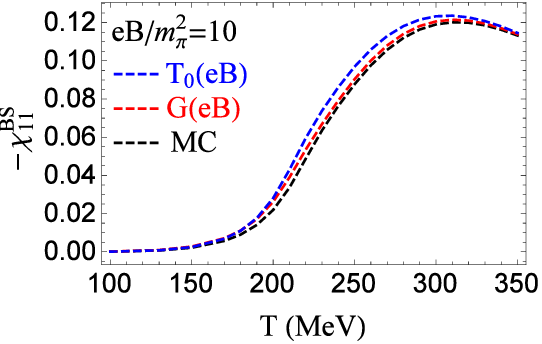}\includegraphics[width=6cm]{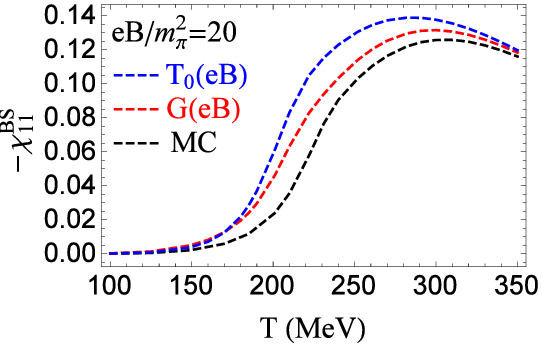}\includegraphics[width=5.8cm]{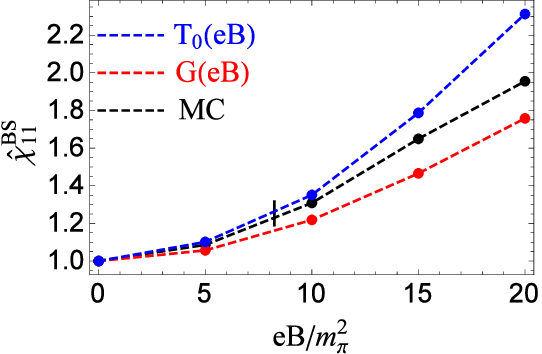}\\
\includegraphics[width=6cm]{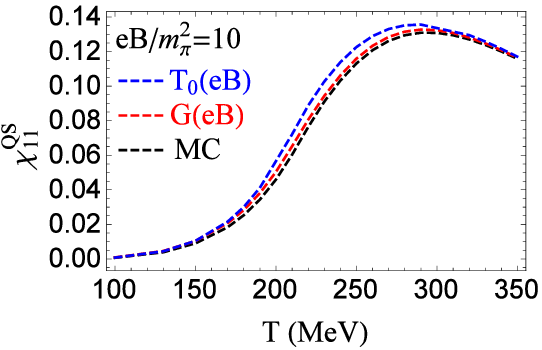}\includegraphics[width=6cm]{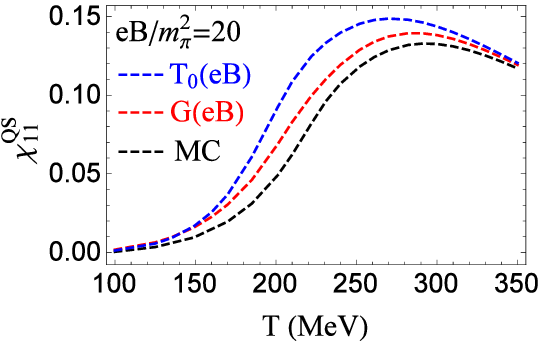}\includegraphics[width=5.8cm]{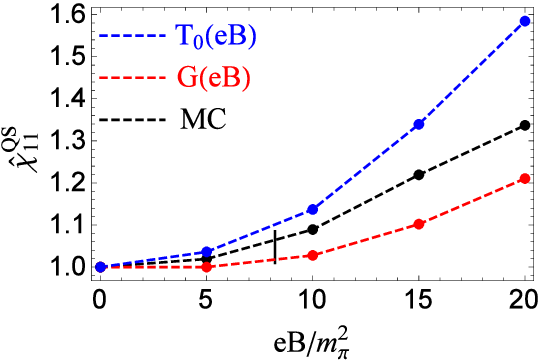}
\end{center}
\caption{(First row) The correlation $\chi^{BQ}_{11}$ as a function of temperature with fixed magnetic field $eB=10m^2_\pi$ (left panel) and $eB=20m^2_\pi$ (middle panel), and the scaled correlation ${\hat{\chi}}^{BQ}_{11}$ (right panel) at $T=T_{pc}^c$ as a function of magnetic field. (Second row) The correlation $-\chi^{BS}_{11}$ as a function of temperature with fixed magnetic field $eB=10m^2_\pi$ (left panel) and $eB=20m^2_\pi$ (middle panel), and the scaled correlation ${\hat{\chi}}^{BS}_{11}$ (right panel) at $T=T_{pc}^c$ as a function of magnetic field. (Third row) The correlation $\chi^{QS}_{11}$ as a function of temperature with fixed magnetic field $eB=10m^2_\pi$ (left panel) and $eB=20m^2_\pi$ (middle panel), and the scaled correlation ${\hat{\chi}}^{QS}_{11}$ (right panel) at $T=T_{pc}^c$ as a function of magnetic field. The black vertical lines in the right panels are the LQCD results~\cite{ding2022,ding2025,dingprl2024}. For all panels, the blue (red) lines are the results in case of considering IMC effect in $T_0(eB)$ $(G(eB))$ scheme, and the black lines are the results without IMC effect.}
\label{figx11imc}
\end{figure*}
\begin{figure*}[htb]
\begin{center}
\includegraphics[width=6cm]{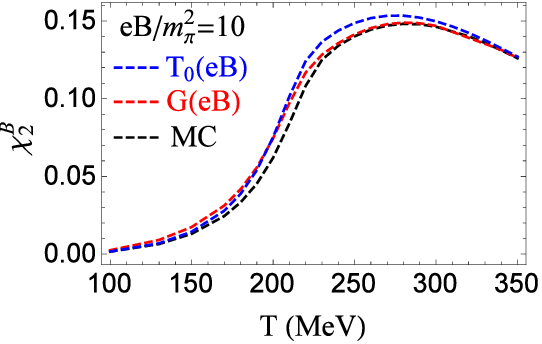}\includegraphics[width=6cm]{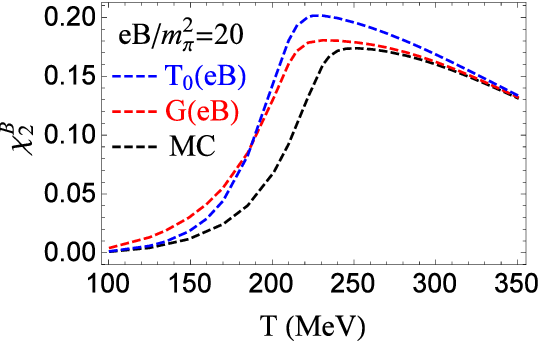}\includegraphics[width=5.8cm]{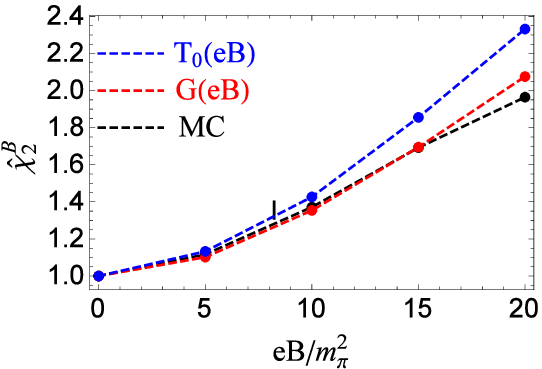}\\
\includegraphics[width=6cm]{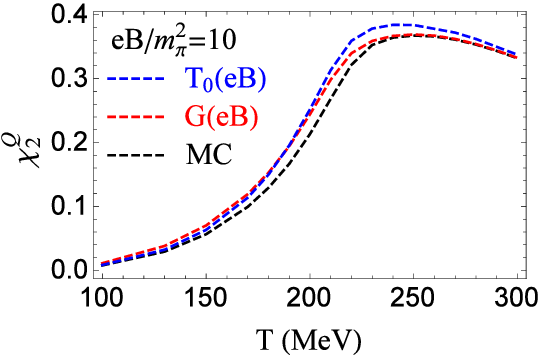}\includegraphics[width=6cm]{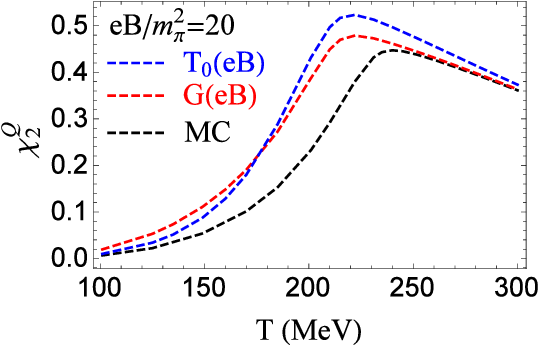}\includegraphics[width=5.9cm]{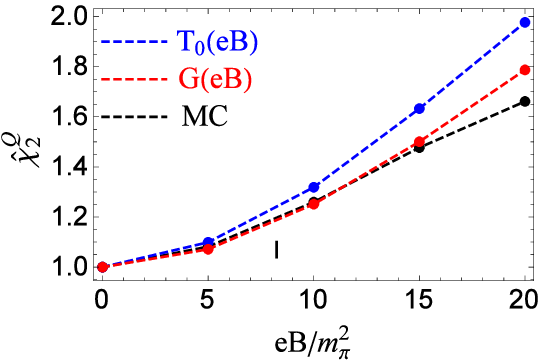}\\
\includegraphics[width=6cm]{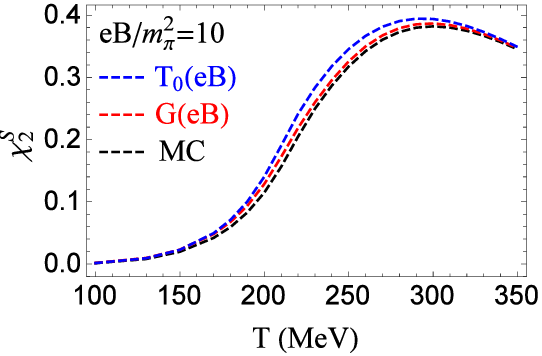}\includegraphics[width=6cm]{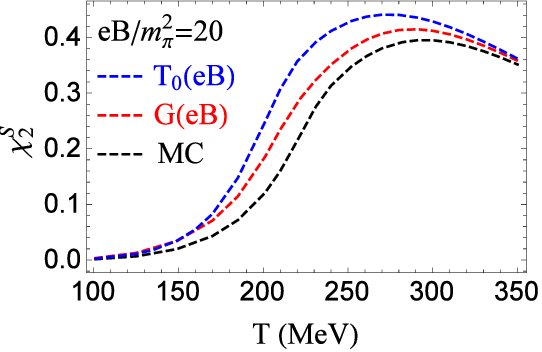}\includegraphics[width=5.9cm]{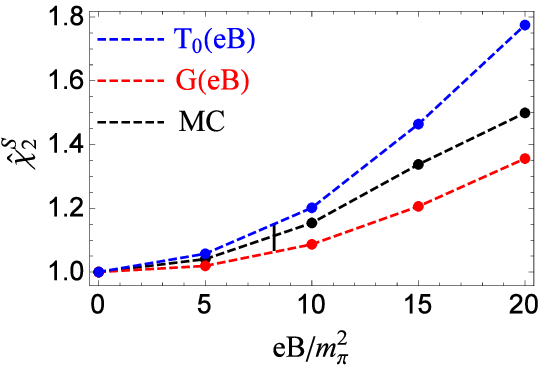}
\end{center}
\caption{(First row) The quadratic fluctuation $\chi^{B}_{2}$ as a function of temperature with fixed magnetic field $eB=10m^2_\pi$ (left panel) and $eB=20m^2_\pi$ (middle panel), and the scaled quadratic fluctuation ${\hat{\chi}}^{B}_{2}$ (right panel) at $T=T_{pc}^c$ as a function of magnetic field. (Second row) The quadratic fluctuation $\chi^{Q}_{2}$ as a function of temperature with fixed magnetic field $eB=10m^2_\pi$ (left panel) and $eB=20m^2_\pi$ (middle panel), and the scaled quadratic fluctuation ${\hat{\chi}}^{Q}_{2}$ (right panel) at $T=T_{pc}^c$ as a function of magnetic field. (Third row) The quadratic fluctuation $\chi^{S}_{2}$ as a function of temperature with fixed magnetic field $eB=10m^2_\pi$ (left panel) and $eB=20m^2_\pi$ (middle panel), and the scaled quadratic fluctuation ${\hat{\chi}}^{S}_{2}$ (right panel) at $T=T_{pc}^c$ as a function of magnetic field. The black vertical lines in the right panels are the LQCD results~\cite{ding2022,ding2025}. For all panels, the blue (red) lines are the results in case of considering IMC effect in $T_0(eB)$ $(G(eB))$ scheme, and the black lines are the results without IMC effect.}
\label{figx2imc}
\end{figure*}
\begin{figure*}[htb]
\begin{center}
\includegraphics[width=6cm]{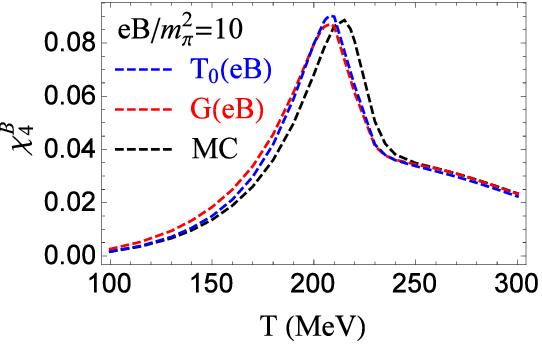}\includegraphics[width=6cm]{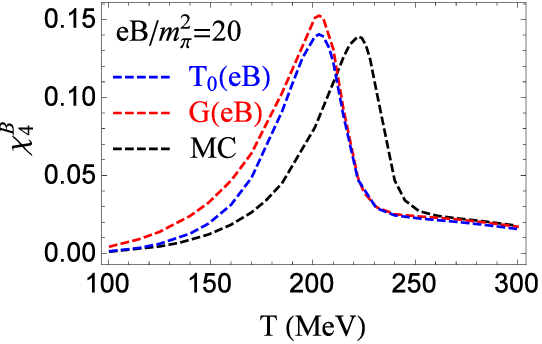}\includegraphics[width=5.8cm]{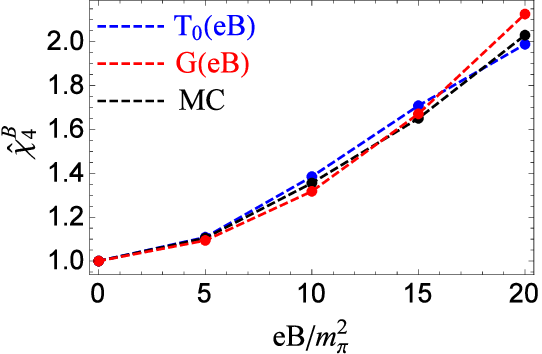}\\
\includegraphics[width=6cm]{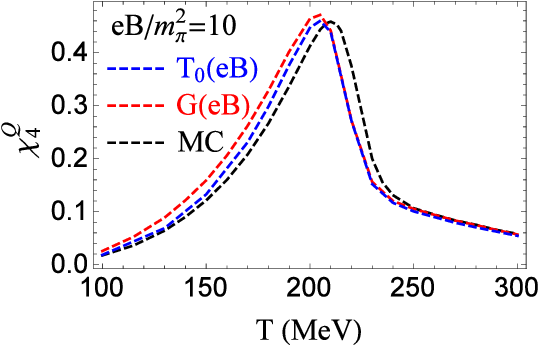}\includegraphics[width=6cm]{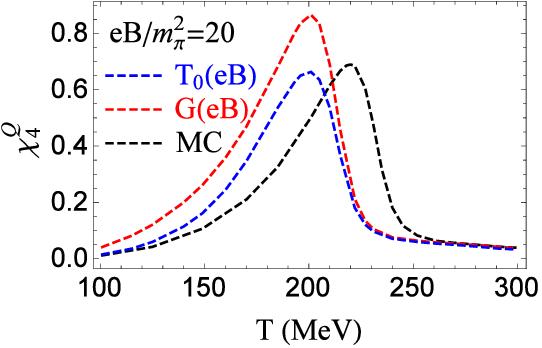}\includegraphics[width=5.9cm]{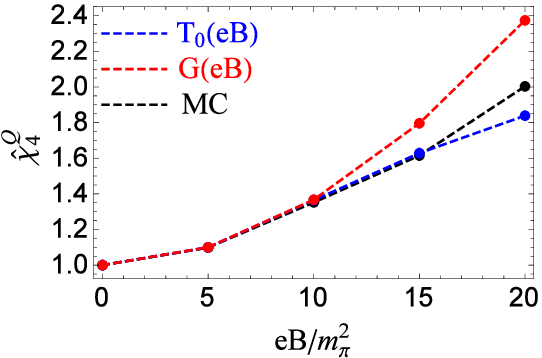}\\
\includegraphics[width=6cm]{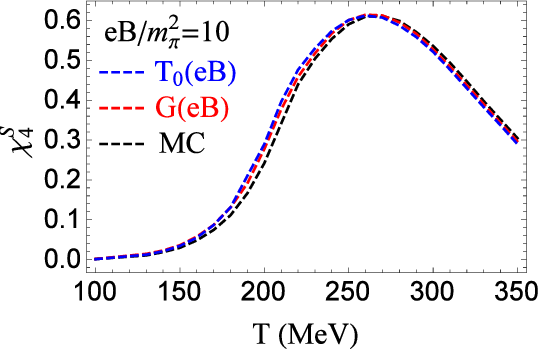}\includegraphics[width=6cm]{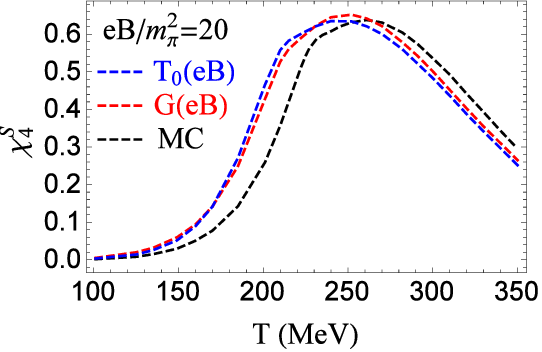}\includegraphics[width=6cm]{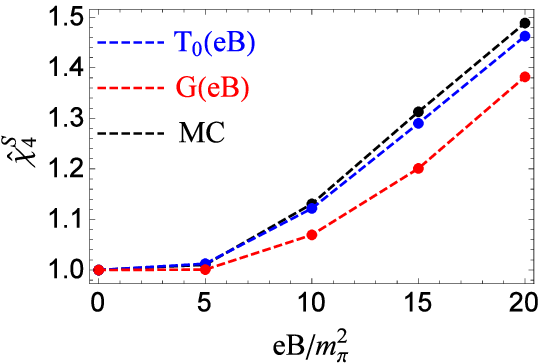}
\end{center}
\caption{(First row) The quartic fluctuation $\chi^{B}_{4}$ as a function of temperature with fixed magnetic field $eB=10m^2_\pi$ (left panel) and $eB=20m^2_\pi$ (middle panel), and the scaled quadratic fluctuation ${\hat{\chi}}^{B}_{4}$ (right panel) at $T=T_{pc}^c$ as a function of magnetic field. (Second row) The quartic fluctuation $\chi^{Q}_{4}$ as a function of temperature with fixed magnetic field $eB=10m^2_\pi$ (left panel) and $eB=20m^2_\pi$ (middle panel), and the scaled quadratic fluctuation ${\hat{\chi}}^{Q}_{4}$ (right panel) at $T=T_{pc}^c$ as a function of magnetic field. (Third row) The quartic fluctuation $\chi^{S}_{4}$ as a function of temperature with fixed magnetic field $eB=10m^2_\pi$ (left panel) and $eB=20m^2_\pi$ (middle panel), and the scaled quadratic fluctuation ${\hat{\chi}}^{S}_{4}$ (right panel) at $T=T_{pc}^c$ as a function of magnetic field. Here, the blue (red) lines are the results in case of considering IMC effect in $T_0(eB)$ $(G(eB))$ scheme, and the black lines are the results without IMC effect.}
\label{figx4imc}
\end{figure*}
\section{Numerical Results}
\label{results}

\subsection{without IMC effect}
Because of the contact interaction in NJL model, the ultraviolet divergence cannot be eliminated through renormalization, and a proper regularization scheme is needed. In this part, we apply the covariant Pauli-Villars regularization~\cite{mao}. By fitting the physical quantities, pion mass $m_{\pi}=138\ \text{MeV}$, pion decay constant $f_{\pi}=93\ \text{MeV}$, kaon mass $m_K=495.7\ \text{MeV}$, $\eta'$ meson mass $m_{\eta\prime}=957.5\ \text{MeV}$ in vacuum, we fix the current masses of light quarks $m_0^{u}=m_0^{d}=5.5\ \text{MeV}$, and obtain the parameters $m_0^s=154.7\ \text{MeV}$, $G\Lambda^2=3.627$, $K\Lambda^5=92.835$, $\Lambda=1101\ \text{MeV}$~\cite{3pnjlmei}. For the Polyakov potential, the parameters are chosen as~\cite{pnjl6} $a_0=6.75$, $a_1=-1.95$, $a_2=2.625$, $a_3=-7.44$, $b_3=0.75$, $b_4=7.5$, and $T_0=270$ MeV.

The typical results of PNJL model with the original parameters determined from the vacuum properties present the catalysis phenomena at finite temperature and vanishing chemical potential. The order parameters, chiral condensates $\sigma_u/\sigma_{u0},\ \sigma_d/\sigma_{d0},\ \sigma_s/\sigma_{s0}$ and Polyakov loop $\Phi$, increase with the magnetic field in the whole temperature region, as shown in Fig.\ref{orderparameterg0t0} first two rows. The last row of Fig.\ref{orderparameterg0t0} depicts the derivative of the order parameters with respect to the temperature $\frac{d \sigma_{ud}}{d T}$ and $\frac{d \Phi}{d T}$ as functions of temperature with fixed magnetic fields. They firstly increase and then decrease with temperature, showing apparent peaks. The location of the peak determines the pseudocritical temperature for chiral restoration and deconfinement phase transitions, respectively, which increase with the magnetic field. The peak value of $\frac{d \sigma_{ud}}{d T}$ and $\frac{d \Phi}{d T}$ determines the strength of chiral restoration and deconfinement phase transitions, respectively, which also increase with the magnetic field. It should be mentioned that the catalysis phenomena of chiral condensates at high temperature and pseudocritical temperature of chiral restoration phase transition under external magnetic field are contrary to LQCD results~\cite{lattice1,lattice2,lattice2sep,lattice4,lattice5,lattice6,lattice7,lattice9,ding2025}, but the increasing strength of chiral restoration phase transition with increasing magnetic field is consistent with LQCD results~\cite{lattice1,lattice2,lattice2sep}.

Figure \ref{figxng0t0} plots quadratic fluctuations $\chi^B_2,\ \chi^Q_2,\ \chi^S_2$ (first row), quartic fluctuations $\chi^B_4,\ \chi^Q_4,\ \chi^S_4$ (second row) and correlations $\chi^{BQ}_{11},\ -\chi^{BS}_{11},\ \chi^{QS}_{11}$ (last row) as functions of temperature with vanishing chemical potential and fixed magnetic field $eB/m^2_{\pi}=0,\ 10,\ 20$. At vanishing magnetic field, $\chi^B_2,\ \chi^Q_2,\ \chi^S_2$ firstly increase and then decrease with temperature, with the wide peak around the chiral restoration and deconfinement phase transitions. Turning on the magnetic field, the peak structure becomes more pronounced. In high temperature region, $\chi^B_2$ and $\chi^Q_2$ have the higher values as magnetic field grows, but in low temperature region they show non-monotonic behavior with increasing magnetic field. $\chi^S_2$ increases with magnetic field in the whole temperature region. For quartic fluctuations $\chi^B_4,\ \chi^Q_4,\ \chi^S_4$, we observe the apparent peak around the chiral restoration and deconfinement phase transitions with and without magnetic field. Their values around the peak are enhanced by the magnetic field. In low temperature region, $\chi^B_4$ and $ \chi^Q_4$ show non-monotonic dependence on the magnetic field, and in high temperature region they decrease under magnetic fields. $\chi^S_4$ is not sensitive to the magnetic field in the temperature region away from the phase transitions. The correlations $\chi^{BQ}_{11},\ -\chi^{BS}_{11},\ \chi^{QS}_{11}$ show peak structure around phase transitions, which are enhanced by the magnetic field. In high temperature region, $\chi^{BQ}_{11}$ and $-\chi^{BS}_{11}$ have the higher values as magnetic field grows, but in low temperature region they show non-monotonic behavior with increasing magnetic field. $\chi^{QS}_{11}$ increases with magnetic field in the whole temperature region. The peak structure of fluctuations and correlations is caused by the phase transition (crossover). When turning on the external magnetic field, the strength of phase transition increases (see Fig.\ref{orderparameterg0t0} last row), and this leads to the enhancement of the fluctuations and correlations around the phase transition. It is noticeable that fluctuations and correlations related with strangeness $(\chi^S_2,\ \chi^S_4,\ -\chi^{BS}_{11},\ \chi^{QS}_{11})$ are less sensitive to the magnetic field than others $(\chi^B_2,\ \chi^Q_2,\ \chi^B_4,\ \chi^Q_4,\ \chi^{BQ}_{11})$, and the peak structure of $\chi^S_2,\ \chi^S_4,\ -\chi^{BS}_{11},\ \chi^{QS}_{11}$ is less apparent than $\chi^B_2,\ \chi^Q_2,\ \chi^B_4,\ \chi^Q_4,\ \chi^{BQ}_{11}$, either. The reason is that all up, down and strange quarks make contribution to $\chi^B_2,\ \chi^Q_2,\ \chi^B_4,\ \chi^Q_4,\ \chi^{BQ}_{11}$, but $\chi^S_2,\ \chi^S_4,\ -\chi^{BS}_{11},\ \chi^{QS}_{11}$ are mainly related to strange quarks, which are much more heavier than up and down quarks. The quadratic fluctuations $\chi^B_2,\ \chi^Q_2,\ \chi^S_2$ and correlation $\chi^{BQ}_{11},\ -\chi^{BS}_{11},\ \chi^{QS}_{11}$ are qualitatively consistent with LQCD simulations~\cite{addding,ding2025}. Comparing with the results in Ref~\cite{dingref35}, where peak structure is observed only in correlation $\chi^{BQ}_{11}$ and quartic fluctuations $\chi^B_4,\ \chi^Q_4,\ \chi^S_4$, the difference is owing to the different regularization schemes, where the regularization is operated in the vacuum term $(T=\mu_f=0)$ of thermodynamical potential in Ref~\cite{dingref35}, but the regularization is applied in both the vacuum and medium term of thermodynamical potential in our current work.

What is the property of correlations and fluctuations along the phase transition line under external magnetic field? As shown in Fig.\ref{figxntpcg0t0}, the scaled correlations ${\hat {\chi}}_{11}^{BQ}=\frac{\chi_{11}^{BQ}(eB,T_{pc}^c(eB))}{\chi_{11}^{BQ}(eB=0,T_{pc}^c(eB=0))}$, ${\hat {\chi}}_{11}^{BS}=\frac{\chi_{11}^{BS}(eB,T_{pc}^c(eB))}{\chi_{11}^{BS}(eB=0,T_{pc}^c(eB=0))}$, ${\hat {\chi}}_{11}^{QS}=\frac{\chi_{11}^{QS}(eB,T_{pc}^c(eB))}{\chi_{11}^{QS}(eB=0,T_{pc}^c(eB=0))}$ and scaled quadratic (quartic) fluctuations ${\hat {\chi}}_{2,4}^{B}=\frac{\chi_{2,4}^{B}(eB,T_{pc}^c(eB))}{\chi_{2,4}^{B}(eB=0,T_{pc}^c(eB=0))}$, ${\hat {\chi}}_{2,4}^{Q}=\frac{\chi_{2,4}^{Q}(eB,T_{pc}^c(eB))}{\chi_{2,4}^{Q}(eB=0,T_{pc}^c(eB=0))}$, ${\hat {\chi}}_{2,4}^{S}=\frac{\chi_{2,4}^{S}(eB,T_{pc}^c(eB))}{\chi_{2,4}^{S}(eB=0,T_{pc}^c(eB=0))}$, at the pseudocritical temperature $T_{pc}^c$ of chiral restoration phase transition are plotted as functions of magnetic field. They are all increasing functions of magnetic field, which is related with the increasing strength of phase transition under external magnetic field, shown in Fig.\ref{orderparameterg0t0} last row. Among the scaled correlations and fluctuations, ${\hat {\chi}}^{BQ}_{11}$ increases fastest, which may serve as a magnetometer of QCD~\cite{dingprl2024,ding2025}. These properties are qualitatively consistent with the LQCD results~\cite{ding2022,ding2025,dingprl2024}, depicted in the vertical lines in Fig.\ref{figxntpcg0t0} left and middle panels. At the quantitative level, ${\hat {\chi}}^{BS}_{11}$, ${\hat {\chi}}^{QS}_{11}$ and ${\hat {\chi}}^{S}_{2}$ are reasonably good, ${\hat {\chi}}^{BQ}_{11}$ and ${\hat {\chi}}^{B}_{2}$ slightly undershoot the LQCD results, and ${\hat {\chi}}^{Q}_{2}$ overshoots the LQCD result.

\subsection{with IMC effect}
From LQCD simulations~\cite{lattice1,lattice2,lattice2sep,lattice4,lattice5,lattice6,lattice7,lattice9}, the IMC phenomenon of chiral symmetry restoration of $u$ and $d$ quarks can be characterized either by the chiral condensates or by the pseudocritical temperature of chiral restoration phase transition. To simulate the IMC effect in the effect model, one approach is to fit the LQCD results of chiral condensates~\cite{geb1,meson,geb3,geb4}, and another approach is to fit the LQCD results of pseudocritical temperature~\cite{bf8,bf9,geb1,mao2pnjl,su3meson4,mao11,t0effectmao}. The two methods give consistent results with each other.

By fitting the LQCD reported decreasing pseudocritical temperature of chiral symmetry restoration $T_{pc}^c(eB)/T_{pc}^c(eB=0)$ under external magnetic field~\cite{lattice1}, we introduce the IMC effect in our three-flavor PNJL model through a magnetic field dependent parameter $G(eB)$ and $T_0(eB)$, respectively, which represent the influence of external magnetic field to the quark-gluon interaction. On one side, the coupling between quarks plays a significant role in determining the spontaneous breaking and restoration of chiral symmetry. Considering the direct interaction between quarks and external magnetic field, a magnetic field dependent coupling $G(eB)$~\cite{bf8,bf9,mao2pnjl,geb1,su3meson4,mao11} is introduced into the PNJL model. On the other side, the interaction between the Polyakov loop and sea quarks may be important for the mechanism of IMC~\cite{lattice9}. A magnetic field dependent Polyakov loop scale parameter $T_0(eB)$~\cite{t0effectmao,t0effect,pnjl3,mao2pnjl} is introduced into the PNJL model to mimic the reaction of the gluon sector to the presence of magnetic fields. As plotted in Fig.\ref{figimcparameter}$, G(eB)$ and $T_0(eB)$ are both monotonic decreasing functions of magnetic field. We have checked that, with our fitted parameter $G(eB)$ or $T_0(eB)$, the increase (decrease) of chiral condensates with magnetic fields at the low (high) temperature, the increase of Polyakov loop with magnetic fields in the whole temperature region and the reduction of pseudocritical temperature of deconfinement phase transition under magnetic fields can be realized. Moreover, the inclusion of IMC effect does not change the strength of chiral restoration phase transition under external magnetic field, see Fig.\ref{figimcparameter} lower panel, where $\frac{d \sigma_{ud}}{d T}$ at the pseudocritical temperature of chiral restoration phase transition increases with magnetic fields and are almost coincident in the cases with and without IMC effect. This indicates that including IMC effect will not lead to qualitative difference in the results of correlations and fluctuations.


In the following part, we use the $G(eB)$ and $T_0(eB)$ scheme to consider the IMC effect, respectively, and make detailed comparison between the results of correlations and fluctuations with and without IMC effect.

Figure \ref{figx11imc} depicts the correlations $\chi^{BQ}_{11}$, $-\chi^{BS}_{11}$, $\chi^{QS}_{11}$ with IMC effect (blue and red lines) and without IMC effect (black lines). At fixed magnetic fields $eB/m^2_\pi=10,\ 20$, when considering the IMC effect with $G(eB)$ and $T_0(eB)$ schemes, the correlations $\chi^{BQ}_{11}$, $-\chi^{BS}_{11}$, $\chi^{QS}_{11}$ increase with temperature and have the peak around the pseudocritical temperatures of chiral restoration and deconfinement phase transitions, which are similar as the case without IMC effect, see the left and middle panels. The value of $\chi^{BQ}_{11}$ with IMC effect are larger (smaller) than those without IMC effect in the low (high) temperature region. The value of $-\chi^{BS}_{11}$ and $\chi^{QS}_{11}$ with IMC effect are larger than those without IMC effect in the whole temperature region. $\chi^{BQ}_{11}$ shows sharper peak than $-\chi^{BS}_{11}$ and $\chi^{QS}_{11}$. In the right panels, we plot the scaled correlations ${\hat {\chi}}_{11}^{BQ}$, ${\hat {\chi}}_{11}^{BS}$, ${\hat {\chi}}_{11}^{QS}$ at the pseudocritical temperature of chiral restoration phase transition as functions of magnetic field with IMC effect (blue and red lines) and without IMC effect (black lines), and LQCD results~\cite{ding2022,ding2025,dingprl2024} are depicted in the black vertical lines for references. ${\hat {\chi}}_{11}^{BQ}$, ${\hat {\chi}}_{11}^{BS}$ and ${\hat {\chi}}_{11}^{QS}$ increase with magnetic field. Different methods to consider IMC effect do not lead to qualitative difference in the results, and only cause some quantitative changes. Including the IMC effect with $G(eB)$ and $T_0(eB)$ schemes, ${\hat {\chi}}^{BQ}_{11}$ increases faster. ${\hat {\chi}}_{11}^{BS}$ and ${\hat {\chi}}_{11}^{QS}$ increase faster (slower) in the $T_0(eB)\ (G(eB))$ scheme.

Figure \ref{figx2imc} plots the quadratic fluctuations $\chi^B_2,\ \chi^Q_2,\ \chi^S_2$ with IMC effect (blue and red lines) and without IMC effect (black lines). At fixed magnetic fields $eB/m^2_\pi=10,\ 20$, when considering the IMC effect with $G(eB)$ and $T_0(eB)$ schemes, $\chi^B_2,\ \chi^Q_2,\ \chi^S_2$ increase with temperature and have the peak around the pseudocritical temperatures of chiral restoration and deconfinement phase transitions, which are similar as the case without IMC effect, see the left and middle panels. Their value with IMC effect are larger than those without IMC effect in the whole temperature region. In the right panels, we plot the scaled quadratic fluctuations ${\hat {\chi}}_2^{B}$, ${\hat {\chi}}_2^{Q}$, ${\hat {\chi}}_2^{S}$ at the pseudocritical temperature of chiral restoration phase transition as functions of magnetic field with IMC effect (blue and red lines) and without IMC effect (black lines), and LQCD results~\cite{ding2022,ding2025} are depicted in the black vertical lines for references. ${\hat {\chi}}_2^{B}$, ${\hat {\chi}}_2^{Q}$ and ${\hat {\chi}}_2^{S}$ increase with magnetic field. Different methods to consider IMC effect do not lead to qualitative difference in the results, and only cause some quantitative changes. Including the IMC effect with $G(eB)$ and $T_0(eB)$ schemes, ${\hat {\chi}}^B_2$ and ${\hat {\chi}}^Q_2$ increase faster than without IMC effect. ${\hat {\chi}}_2^{S}$ increases faster in the $T_0(eB)$ scheme and increases slower in the $G(eB)$ scheme.

Figure \ref{figx4imc} shows the quartic fluctuations $\chi^{B}_{4}$, $\chi^{Q}_{4}$, $\chi^{S}_{4}$ with IMC effect (blue and red lines) and without IMC effect (black lines). At fixed magnetic fields $eB/m^2_\pi=10,\ 20$, when considering the IMC effect with $G(eB)$ and $T_0(eB)$ schemes, $\chi^{B}_{4}$, $\chi^{Q}_{4}$ and $\chi^{S}_{4}$ increase with temperature and have the peak around the pseudocritical temperatures of chiral restoration and deconfinement phase transitions, which are similar as the case without IMC effect, see the left and middle panels. Their values with IMC effect are larger (smaller) than those without IMC effect in the low (high) temperature region. $\chi^{B}_{4}$ and $\chi^{Q}_{4}$ show sharper peak than $\chi^{S}_{4}$. In the right panels, we plot the scaled quartic fluctuations ${\hat {\chi}}_4^{B}$, ${\hat {\chi}}_4^{Q}$, ${\hat {\chi}}_4^{S}$ at the pseudocritical temperature of chiral restoration phase transition as functions of magnetic field with IMC effect (blue and red lines) and without IMC effect (black lines). They increase with magnetic field. Different methods to consider IMC effect do not lead to qualitative difference in the results, and only cause some quantitative changes. Including the IMC effect does not alter ${\hat {\chi}}_4^{B}$ too much. ${\hat {\chi}}_4^{Q}$ increases faster (slower) in the $G(eB)\ (T_0(eB))$ scheme. Including the IMC effect with $G(eB)$ and $T_0(eB)$ schemes, ${\hat {\chi}}_4^{S}$ increases slower than without IMC effect.

\section{summary}
\label{summary}
The correlations $\chi^{BQ}_{11},\ \chi^{BS}_{11} ,\ \chi^{QS}_{11}$ and fluctuations $\chi^B_{2,4},\ \chi^Q_{2,4},\ \chi^S_{2,4}$ at finite temperature and magnetic field are investigated in frame of a three-flavor PNJL model. The IMC effect is introduced into the PNJL model by two methods, the magnetic field dependent coupling between quarks $G(eB)$, and magnetic field dependent interaction between quarks and Polyakov loop $T_0(eB)$. We make comparison of the results in the cases with and without IMC effect.

The PNJL model with IMC effect recovers the decreasing pseudocritical temperature of phase transitions under external magnetic field and the properties of order parameters reported in LQCD simulations. But the increasing strength of chiral restoration phase transition under external magnetic field is obtained in the cases with and without IMC effect. Therefore, the inclusion of IMC effect does not lead to qualitative difference in the results of correlations and fluctuations, and only causes some quantitative changes.

The correlations $\chi^{BQ}_{11},\ \chi^{BS}_{11} ,\ \chi^{QS}_{11}$ and fluctuations $\chi^B_{2,4},\ \chi^Q_{2,4},\ \chi^S_{2,4}$ show similar properties at finite temperature and magnetic field. With and without magnetic fields, the correlations and fluctuations first increase with temperature, and then show the peak around the pseudocritical temperatures of chiral restoration and deconfinement phase transitions. The peak structure in $\chi^{BQ}_{11}$, $\chi^B_{4}$ and $\chi^Q_{4}$ is much more apparent than in other correlations and fluctuations. To demonstrate the correlations and fluctuations along the phase transition line under external magnetic field, we calculate the scaled correlations ${\hat {\chi}}_{11}^{BQ}$, ${\hat {\chi}}_{11}^{BS}$, ${\hat {\chi}}_{11}^{QS}$ and scaled fluctuations ${\hat {\chi}}_{2,4}^{B}$, ${\hat {\chi}}_{2,4}^{Q}$, ${\hat {\chi}}_{2,4}^{S}$ at the pseudocritical temperature of chiral restoration phase transition. They increase with magnetic fields, which is caused by the increase of phase transition strength under external magnetic field. Among them, ${\hat {\chi}}_{11}^{BQ}$ increases fastest, which may serve as the magnetometer of QCD. These properties of correlations and quadratic fluctuations are qualitatively consistent with the LQCD results.\\ 

\noindent {\bf Acknowledgement:} The work is supported by the NSFC grant 12275204.


\begin{thebibliography}{99}

\bibitem{review1} F.Preis, A.Rebhan and A.Schmitt, Lect. Notes Phys. {\bf 871}, 51(2013).
\bibitem{review2} R.Gatto and M.Ruggieri, Lect. Notes Phys. {\bf 871}, 87(2013).
\bibitem{review4} V.A.Miransky and I.A.Shovkovy, Phys. Rep. {\bf 576}, 1(2015).
\bibitem{review5} J.O.Anderson and W.R.Naylor, Rev. Mod. Phys. {\bf 88}, 025001(2016).
\bibitem{review0} G.Q.Cao, Eur. Phys. J. {\bf A57}, 264(2021).

\bibitem{review3} M.D'Elia, Lect. Notes Phys. {\bf 871}, 181(2013).
\bibitem{lattice1} G.S.Bali, F.Bruckmann, G.Endr$\ddot{o}$di, Z.Fodor, S.D.Katz, S.Krieg, A.Schaefer and K.K.Szabo, J. High Energy Phys. {\bf 02}, 044(2012).
\bibitem{lattice2} G.S.Bali, F.Bruckmann, G.Endr$\ddot{o}$di, Z.Fodor, S.D.Katz and A.Schaefer, Phys. Rev. {\bf D86}, 071502(2012).
\bibitem{lattice2sep} G.S.Bali, F.Bruckmann, G.Endr$\ddot{o}$di, Z.Fodor, S.D.Katz and A.Schaefer, J. High Energy Phys. {\bf 08}, 177(2014).

\bibitem{lattice9} F.Bruckmann, G.Endr$\ddot{o}$di and T.G.Kovacs, J. High Energy Phys. {\bf 04}, 112(2013).
\bibitem{lattice4} V.G.Bornyakov, P.V.Buividovich, N.Cundy, O.A.Kochetkov and A.Schaefer, Phys. Rev. {\bf D90}, 034501(2014).
\bibitem{lattice5} G.Endr$\ddot{o}$di, J. High Energy Phys. {\bf 07}, 173(2015).
\bibitem{lattice6} G.Endr$\ddot{o}$di, M.Giordano, S.D.Katz, T.G.Kovacs and F.Pittler, J. High Energy Phys. {\bf 07}, 009(2019).
\bibitem{lattice7} H.T.Ding, S.T.Li, J.H.Liu and X.D.Wang, Phys. Rev. {\bf D 105}, 034514(2022).
\bibitem{ding2025} H.T.Ding, J.B.Gu, A.Kumar and S.T.Li, arXiv: 2503.18467.

\bibitem{lattice8} M.D'Elia, F.Manigrasso, F.Negro and F.Sanfilippo, Phys. Rev. {\bf D98}, 054509(2018).


\bibitem{fukushima} K.Fukushima and Y.Hidaka, Phys. Rev. Lett {\bf 110}, 031601(2013).
\bibitem{mao} S.J.Mao, Phys. Lett. {\bf B758}, 195(2016).
\bibitem{maosep1}  S.J.Mao, Phys. Rev. {\bf D94}, 036007(2016).
\bibitem{maosep2}  S.J.Mao, Phys. Rev. {\bf D97}, 011501(R)(2018).
\bibitem{maosep3}  S.J.Mao, Chin. Phys. {\bf C45}, 021004(2021).
\bibitem{maosep4}  S.J.Mao, Phys. Rev. {\bf D106}, 034018(2022).
\bibitem{kamikado}  K.Kamikado and T.Kanazawa, J. High Energy Phys. {\bf 03}, 009(2014).
\bibitem{bf1} J.Y.Chao, P.C.Chu and M.Huang, Phys. Rev. {\bf D88}, 054009(2013).

\bibitem{bf13} J.Braun, W.A.Mian and S.Rechenberger, Phys. Lett. {\bf B755}, 265(2016).
\bibitem{bf2} N.Mueller and J.M.Pawlowski,  Phys. Rev. {\bf D91}, 116010(2015).
\bibitem{bf3} T.Kojo and N.Su, Phys. Lett. {\bf B720}, 192(2013).
\bibitem{bf5} A.Ayala, M.Loewe, A.J.Mizher and R.Zamora, Phys. Rev. {\bf D90}, 036001(2014).
\bibitem{bf51}A.Ayala, L.A.Hernandez, A.J.Mizher, J.C.Rojas and C.Villavicencio, Phys. Rev. {\bf D89}, 116017(2014).
\bibitem{bf52}A.Ayala, C.A.Dominguez, L.A.Hernandez, M.Loewe and R.Zamora, Phys. Rev. {\bf D92}, 096011(2015).
\bibitem{bf8} R.L.S.Farias, K.P.Gomes, G.Krein, and M.B.Pinto, Phys. Rev. {\bf C90}, 025203(2014).
\bibitem{bf9} M.Ferreira, P.Costa, O.Lourenco, T.Frederico, and C.Provid$\hat e$ncia, Phys. Rev. {\bf D89}, 116011(2014).
\bibitem{bf11} F.Preis, A.Rebhan and A.Schmitt, J. High Energy Phys. {\bf 1103}, 033(2011).
\bibitem{db1} E.S.Fraga and A.J.Mizher, Phys. Rev. {\bf D78}, 025016(2008).
\bibitem{db1sep} E.S.Fraga and A.J.Mizher, Nucl. Phys. {\bf A820}, 103C(2009).
\bibitem{db2} K.Fukushima, M.Ruggieri and R.Gatto, Phys. Rev. {\bf D81}, 114031(2010).
\bibitem{db3} C.V.Johnson and A.Kundu, J. High Energy Phys. {\bf 12}, 053(2008).
\bibitem{db5} V.Skokov, Phys. Rev. {\bf D85}, 034026(2012).
\bibitem{db6} E.S.Fraga, J.Noronha and L.F.Palhares, Phys. Rev. {\bf D87}, 114014(2013).
\bibitem{pnjl1} R.Gatto and M.Ruggieri, Phys. Rev. {\bf D82}, 054027(2010).
\bibitem{pnjl1sep} R.Gatto and M.Ruggieri, Phys. Rev. {\bf D83}, 034016(2011).
\bibitem{pnjl2} M.Ferreira, P.Costa and C.Provid$\hat e$ncia, Phys. Rev. {\bf D89}, 036006(2014).
\bibitem{pnjl3} M.Ferreira, P.Costa, D.P.Menezes, C.Provid$\hat e$ncia and N.N.Scoccola, Phys. Rev. {\bf D89}, 016002(2014).
\bibitem{pnjl4} P.Costa,  M.Ferreira,  H.Hansen, D.P.Menezes and C.Provid$\hat e$ncia, Phys. Rev. {\bf D89}, 056013(2014).
\bibitem{pqm} A.J.Mizher, M.N.Chernodub and E.S.Fraga, Phys. Rev. {\bf D82}, 105016(2010).
\bibitem{t0effect} E.S.Fraga, B.W. Mintz and J.Schaffner-Bielich, Phys. Lett {\bf B731}, 154-158(2014).
\bibitem{ferr1} E.J.Ferrer, V.de la Incera, I.Portillo and M.Quiroz, Phys. Rev. {\bf D89}, 085034(2014).
\bibitem{ferr2} E.J.Ferrer, V.de la Incera, and X.J.Wen, Phys. Rev. {\bf D91}, 054006(2015).
\bibitem{meimao1} J.Mei and S.J.Mao, Phys. Rev. {\bf D102}, 114035(2020).
\bibitem{mhuang} K.Xu, J.Y.Chao and M.Huang, Phys. Rev. {\bf D103}, 076015(2021).
\bibitem{meihuangmao} J.Mei, R.Wen, S.J.Mao, M.Huang and K.Xu, Phys. Rev. {\bf D110}, 034024(2024).
\bibitem{t0effectmao} S.J.Mao, Phys. Rev. {\bf D110}, 054002(2024).

\bibitem{efield1} G.Cao and X.G.Huang, Phys. Rev. {\bf D93}, 016007(2016).
\bibitem{efield2} W.R.Tavares, R.L.S.Farias and S.S.Avancini, Phys. Rev. {\bf D101}, 016017(2020).
\bibitem{efield3} W.R.Tavares, S.S.Avancini and R.L.S.Farias, Phys. Rev. {\bf D108}, 016017(2023).
\bibitem{efield4} G.Endr$\ddot{o}$di and G.Marko, Phys. Rev. {\bf D109}, 034506(2024).


\bibitem{xwork24} H.T.Ding, F.Karsch and S.Mukherjee, Int. J. Mod. Phys. {\bf E24}, 1530007(2015).
\bibitem{xwork25} W.J.Fu, Commun. Theor. Phys. {\bf 74}, 097304 (2022).
\bibitem{xwork26} X.Luo and N.Xu, Nucl. Sci. Tech. {\bf 28}, 112(2017).
\bibitem{xwork27} A.Pandav, D.Mallic and B.Mohanty, Prog. Part. Nucl. Phys. {\bf 125}, 103960(2022).
\bibitem{xwork28} A.Rustanmov, EPJ Web Conf. {\bf 276}, 01007(2023).
\bibitem{xwork29} T. Nonaka, Acta Phys. Pol. B Proc. Suppl. {\bf 16}, 1-A14(2023).
\bibitem{xwork30} H.S.Ko(STAR  Collaboration), Acta Phys. Pol. B Proc. Suppl. {\bf 16}, 1-A87(2023).


\bibitem{addding}H.T.Ding, S.T.Li, Q.Shi, and X.D.Wang, Eur. Phys. J. {\bf A57}, 202(2021).
\bibitem{ding2022} H.T.Ding, S.T.Li, J.H.Liu and X.D.Wang, arXiv:2208.07285.
\bibitem{dingprl2024}H.T.Ding, J.B.Gu, A.Kumar, S.T.Li and J.H.Liu, Phys. Rev. Lett. {\bf 132}, 201903(2024).


\bibitem{dingref31}K.Fukushima and Y.Hidaka, Phys. Rev. Lett. {\bf 117}, 102301(2016).
\bibitem{dingref33}A.Bhattacharyya, S.K.Ghosh, R.Ray, and S.Samanta, Eur. Phys. Lett. {\bf 115}, 62003(2016).
\bibitem{dingref34}G.Kadam, S.Pal, and A.Bhattacharyya, J. Phys. G {\bf 47}, 125106(2020).
\bibitem{dingref35}W.J.Fu, Phys. Rev. {\bf D88}, 014009(2013).
\bibitem{dingref36}N.Chahal, S.Dutt, and A.Kumar, Phys. Rev. {\bf C107}, 045203(2023).
\bibitem{mao2pnjl} S.J. Mao, Chin. Phys. C (accepted); arXiv:2410.10217.

\bibitem{geb1} H.Liu, L.Yu, M.Chernodub and M.Huang, Phys. Rev. {\bf D94}, 113006(2016).
\bibitem{su3meson4} S.S.Avancini, M.Coppola, N.N.Scoccola, and J.C.Sodr\'{e}, Phys. Rev. {\bf D104}, 094040(2021).
\bibitem{mao11} S.J.Mao and Y.M.Tian, Phys. Rev. {\bf D106}, 094017(2022).



\bibitem{pnjl5} P.N.Meisinger and M.C.Ogilvie, Phys. Lett. {\bf B379}, 163(1996).
\bibitem{pnjl6} P.N.Meisinger, T.R.Miller and M.C.Ogilvie, Phys. Rev. {\bf D65}, 034009(2002).
\bibitem{pnjl7} K.Fukushima, Phys. Lett. {\bf B591}, 277(2004).
\bibitem{pnjl8} A.Mocsy, F.Sannino, and K.Tuominen, Phys. Rev. Lett. {\bf 92}, 182302(2004).
\bibitem{pnjl9} E.Megias, E.Ruiz Arriola, and L.L.Salcedo, Phys. Rev. {\bf D74}, 065005(2006).
\bibitem{pnjl10} C.Ratti, M.A.Thaler, and W.Weise, Phys. Rev. {\bf D73}, 014019(2006).
\bibitem{pnjl12} S.K.Ghosh, T.K.Mukherjee, M.G.Mustafa and R.Ray, Phys. Rev. {\bf D73}, 114007(2006).

\bibitem{tHooft1} T.Kunihiro and T.Hatsuda, Phys. Lett. {\bf B206}, 385(1988).
\bibitem{tHooft2} V.Bernard, R.L.Jaffe, and U.G.Meissner, Nucl. Phys. {\bf B308}, 753(1988).
\bibitem{tHooft3} H.Reinhardt and R.Alkofer, Phys. Lett. {\bf B207}, 482(1988).
\bibitem{tHooft4} G.'t Hooft, Phys. Rev. {\bf D14}, 3432(1976).
\bibitem{tHooft5} G.'t Hooft, Phys. Rept. {\bf 142}, 357(1986).

\bibitem{3njlrehberg} P.Rehberg, S.P.Klevansky, and J.Huefner, Phys. Rev. {\bf C53}, 410(1996).

\bibitem{3pnjlmei}J.Mei, T.Xia, and S.J.Mao, Phys. Rev. {\bf D107}, 074018(2023); {\bf D110}, 119901(E)(2024).


%
%


\bibitem{geb3} A.Ayala, C.A.Dominguez, L.A.Hern$\acute a$ndez, M.Loewe, A.Raya, J.C.Rojas and C.Villavicenico, Phys. Rev. {\bf D94}, 054019 (2016).
\bibitem{geb4} R.L.S.Farias, V.S.Timoteo, S.S.Avancini, M.B.Pinto and G.Klein, Eur. Phys. J. {\bf A53}, 101(2017).
\bibitem{meson} S.Avancini, R.Farias, M.Pinto, W.Travres and V.Tim$\acute{o}$teo, Phys. Lett. {\bf B767}, 247(2017).




\end{thebibliography}
\end{document}